\documentclass[aps,pra,twocolumn,notitlepage]{revtex4-1}

\usepackage{graphicx}
\begin{document}

\title{Quantum interference effects determine contextual realities}

\author{Holger F. Hofmann}
\email{hofmann@hiroshima-u.ac.jp}
\affiliation{
Graduate School of Advanced Science and Engineering, Hiroshima University,
Kagamiyama 1-3-1, Higashi Hiroshima 739-8530, Japan}

\begin{abstract}
Quantum mechanics describes the relation between different measurement contexts in terms of superpositions of the potential measurement outcomes. This relation between measurement contexts makes it impossible to determine context independent realities. Here, I illustrate the problem using three path interferences that implement the three box paradox and show that the representation of the final measurement outcome as a superposition of intermediate paths describes well-defined coherences between seemingly empty paths. 
\end{abstract}

\keywords{quantum measurement, quantum contextuality, quantum interference}

\maketitle

\section{Introduction}

Quantum contextuality is commonly understood in terms of the Kochen-Spekker theorem, which poses the problem in terms of context dependent hidden variable theories \cite{KS}. However, contextuality can also be understood as a fundamental feature of quantum statistics, corresponding to the appearance of negative quasi-probabilities in joint probabilities of measurements that cannot be performed jointly \cite{Spe05,Pus14,Sch18}. It is therefore not surprising that many of the paradoxes introduced by Yakir Aharonov and his coworkers based on weak measurements and post-selection provide clear evidence for quantum contextuality \cite{Aha88,Lei05,Tol07}. Of particular interest is the well-known 3-box paradox, which is the most simple example of a paradox associated with the appearance of negative quasi-probabilities in weak measurements \cite{Res04,Hof15}. This paradox is easily observed in a simple three-path interferometer, where the logic of classical three path interference can be applied to discuss the meaning of contextuality and its fundamental relation with the concept of quantum superpositions \cite{Ji23}. In the following, I will explain why a conventional three-path interference experiment can reveal paradoxical aspects of quantum contextuality and identify the specific quantum coherence responsible for the appearance of negative quasi-probabilities and the associated contextuality. In particular, it will be shown that the detection of a single particle in a specific output port effectively separates the input state coherences from their paths in such a way as to define coherences between paths that are otherwise empty. Such ``ghost coherences'' describe the contextual relations between different measurements and ensure that the realities of one measurement context cannot be combined with hypothetical realities of intermediate measurements when such measurements are impossible to perform. It is important to recognize that measurement interactions have a well-defined impact on the nature of objective realities, since the effects of a system cannot be separated from the dynamics of quantum measurements \cite{Mat23}. It is therefore necessary to consider the possibility that physical properties can have values other than their eigenvalues in interactions that are too weak to break the deterministic relation with future measurements \cite{Hof21,Lem22,Hof23}.

\section{The paradox of three path interference}

\begin{figure}
\begin{picture}(240,120)
%%\put(0,0){\framebox(240,120){}}
\put(0,0){\makebox(240,120){\vspace{-2cm}
\scalebox{0.6}[0.6]{
\includegraphics{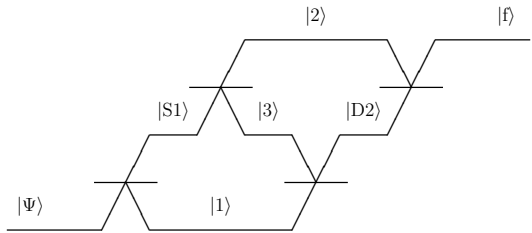}}}}
\end{picture}

\caption{\label{fig1}  
Three-path interferometer illustrating the three-box paradox. The reflectivities of the first and fourth beam splitter are 1/3, and the reflectivities of the second and third beam splitter are 1/2, respectively. Photons enter in $\mid \psi \rangle$ and are detected in $\mid f \rangle$. Due to destructive interference, no photon passes through $\mid D_2 \rangle$. It seems that all of the photons passing from $\mid \psi \rangle$ to $\mid f \rangle$ pass through $\mid S_1 \rangle$. However, the transmission from $\mid S_1 \rangle$ to $\mid f \rangle$ is zero, since $\langle f \mid S_1 \rangle=0$. 
}

\end{figure}

As its name implies, the three-box paradox is usually formulated in the abstract, as if one could look at quantum particles as classical objects. However, all experimental realizations necessarily implement it as a three-path interferometer. It is therefore better to discuss the paradox itself in terms of its most obvious physical implementation using a single photon and a sequence of beam splitters. Fig. \ref{fig1} shows the setup implementing the three-box paradox. The state $\mid \psi \rangle$ of the photon is defined by its input port in a three path interferometer, and the measurement outcome $\mid f \rangle$ is defined by the output port in which the photon is detected. The question is whether the specific combination of $\mid \psi \rangle$ and $\mid f \rangle$ puts any logical constraints on the path taken by the particle between $\mid \psi \rangle$ and $\mid f \rangle$. In the forward direction in time, the wavefunction of the photon is split at the first beam splitter, which has a reflectivity of $1/3$,
\begin{equation}
\mid \psi \rangle = \sqrt{\frac{2}{3}} \mid S1 \rangle + \sqrt{\frac{1}{3}} \mid 1 \rangle. 
\end{equation}
The amplitude in $\mid S1 \rangle$ is then split into an equal superposition of $\mid 2 \rangle$ and $\mid 3 \rangle$, and the amplitude in $\mid 3 \rangle$ then interferes with the amplitude in $\mid 1 \rangle$. Since both amplitudes are equal, they interfere destructively in the direction of $\mid D2 \rangle$, reflecting the fact that the initial state $\mid \psi \rangle$ is orthogonal to $\mid D2 \rangle$,
\begin{equation}
\langle D2 \mid \psi \rangle = 0.
\end{equation}
The path $\mid D2 \rangle$ is therefore empty, independent of the future detection of the photon. The only path that leads to $\mid f \rangle$ is the path through $\mid S1 \rangle$. One might therefore think that detection of the photon in $\mid f \rangle$ ensures passage of the photon through $\mid S1 \rangle$. However, this seems to be impossible, because the transmission from $\mid S1 \rangle$ to $\mid f \rangle$ is given by 
\begin{equation}
\langle f \mid S1 \rangle = 0.
\end{equation}
The photon cannot be found in both $\mid S1 \rangle$ and $\mid f \rangle$ under any circumstances. If the photon travels from $\mid \psi \rangle$ and $\mid f \rangle$, both $\mid S1 \rangle$ and $\langle D2 \mid \psi \rangle$ should be empty.

\section{Evidence from weak interactions}

\begin{figure}
\begin{picture}(240,120)
%%\put(0,0){\framebox(240,120){}}
\put(0,0){\makebox(240,120){\vspace{-2cm}
\scalebox{0.6}[0.6]{
\includegraphics{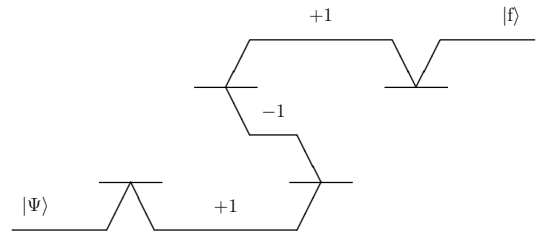}}}}
\end{picture}

\caption{\label{fig2}  
Weak values of path presence. Particles traveling from $\mid \psi \rangle$ to $\mid f \rangle$ have a negative presence in path $\mid 3 \rangle$, indicating that individual particles detected in the context $\mid f \rangle$ are physically delocalized in the paths $\mid 1 \rangle$, $\mid 2 \rangle$ and $\mid 3 \rangle$.
}
\end{figure}

Weak measurements can be used to confirm that the paths $\mid S1 \rangle$ and $\langle D2 \mid \psi \rangle$ are indeed empty. According to the weak values of the intermediate path presence given by the path projectors, all of the photons traveling from $\mid \psi \rangle$ to $\mid f \rangle$ are reflected into path $\mid 1 \rangle$ at the first beam splitter,
\begin{equation}
\frac{\langle f \mid 1 \rangle \langle 1 \mid \psi \rangle}{\langle f \mid \psi \rangle} = +1. 
\end{equation}
When this conditional photon current arrives at the third beam splitter, there is no way for it to continue. Weak values solve this problem by indicating that there is a negative current in path $\mid 3 \rangle$,
\begin{equation}
\frac{\langle f \mid 3 \rangle \langle 3 \mid \psi \rangle}{\langle f \mid \psi \rangle} = -1. 
\end{equation}
This negative conditional current originated from the second beam splitter, where a current of zero was split into the negative current in $\mid 3 \rangle$ and a positive current in $\mid 2 \rangle$,
\begin{equation}
\frac{\langle f \mid 2 \rangle \langle 2 \mid \psi \rangle}{\langle f \mid \psi \rangle} = +1. 
\end{equation}
Weak values thus provide a consistent explanation of photon currents between $\mid \psi \rangle$ and $\mid f \rangle$ by assigning a negative current to the central path $\mid 3 \rangle$. Fig. \ref{fig2} illustrates the propagation of the particles through the interferometer.  

Obviously, there is no simple way in which the propagation pattern determined by the weak values can be interpreted. However, there seems to be strong evidence that particles observed in contexts described by interferences between different paths can be delocalized over the paths that are not directly detected \cite{Hof21,Lem22,Hof23}. Importantly, this possibility might explain why the wave-particle dualism fails to provide a satisfactory interpretation of quantum mechanics. Since the present discussion concerns a simple single particle interference experiment, one might ask what its implications are for classical waves. If the experiment is done with strong laser light, the output intensity observed in $\mid f \rangle$ could be traced through the interferometer by small modifications of its polarization, similar to the methods proposed in \cite{Hof21,Lem22,Hof23}. What would the negative results in $\mid 3 \rangle$ mean then? 

Surprisingly, there is no simple answer to this question. Wave propagation conserves energy, but the origin of a specific part of the energy cannot be determined. Energy has no identity, and it cannot be objectified in the same way that a particle or a wave amplitude can be objectified. The difference between quantum interference and classical wave interference is that the causality of interactions refers to the presence of the particle in a specific path, while the causality of wave propagation refers to the amplitudes as the fundamental physical object. In wave optics, it is the energy that has a somewhat elusive character. It may actually be reasonable to consider a negative energy current in an interferometer, but it is not possible to separate the part of the energy that arrives at $\mid f \rangle$ from the energy that exits at other ports of the interferometer. Ultimately, it is the observation of individual measurement outcomes that distinguishes quantum interference from classical interference. 

\section{Coherences between empty paths}

The analogy with classical wave coherence may have its flaws, but it is still very helpful with regard to the fundamental oddity of the appearance of negative and positive currents from paths that should have been be empty. Quantum interference is fully deterministic since the quantum coherence that determines the output statistics is already present in the input. For the second beam splitter, we can identify the output paths with specific superpositions of the input paths. If we introduce the empty input path $\mid D1 \rangle$, the relation between output paths and input coherence can be expressed by the operator relation
\begin{equation}
\mid 2 \rangle \langle 2 \mid - \mid 3 \rangle \langle 3 \mid = \mid D1 \rangle \langle S1 \mid + \mid S1 \rangle \langle D1 \mid.
\end{equation}
Likewise, the input paths of the third beam splitter are converted to an output coherence between the empty paths $\mid D2 \rangle$ and $\mid S2 \rangle$,
\begin{equation}
\mid D2 \rangle \langle S2 \mid + \mid S2 \rangle \langle D2 \mid = \mid 1 \rangle \langle 1 \mid - \mid 3 \rangle \langle 3 \mid.
\end{equation}
Most misunderstandings regarding the propagation of quantum particles arise because coherences are treated as esoteric properties, even though they have an equal claim to reality as the position of the particle. We tend to be biased towards a completely static picture of reality that does not consider the dynamic relations between different contexts described by the operator formalism. 

As I have pointed out in several previous works \cite{WV1,NJP11,NJP12}, the quantum statistics of weak values can be interpreted as a measurement based decomposition of the density matrix. This decomposition can separate coherences between paths from the projectors onto these paths, resulting in an effective coherence between paths that appear to be empty. In the present case, the pre- and post-selected state can be expressed as
\begin{eqnarray}
\frac{\mid \psi \rangle \langle f \mid}{\langle f \mid \psi \rangle} &=& \mid 1 \rangle \langle 1 \mid + \sqrt{2}
\mid S1 \rangle \langle 1 \mid 
\nonumber \\ &&
+ \sqrt{2} \mid 1 \rangle \langle D1 \mid + 2 \mid S1 \rangle \langle D1 \mid.
\end{eqnarray}
This representation describes the situation between the first and the second beam splitter, where the photon appears in path $\mid 1 \rangle$, but there is already a coherence between paths $\mid S1 \rangle$ and $\mid D1 \rangle$. This coherence is an essential aspect describing the relations between different measurement contexts. The weak value of 1 in path $\mid 1 \rangle$ represents the identification of the particle path based on the combination of separate statements about $\mid \psi \rangle$ and $\mid f \rangle$. However, quantum mechanics makes such combination problematic, as expressed by the coherences between the empty paths $\mid S1 \rangle$ and $\mid D1 \rangle$. In the three-box paradox, the problem can be seen when we consider path $\mid 2 \rangle$. We can do this by representing the same pre- and post-selected state between the third and the fourth beam splitter, where  
\begin{eqnarray}
\frac{\mid \psi \rangle \langle f \mid}{\langle f \mid \psi \rangle} &=& \mid 2 \rangle \langle 2 \mid + \sqrt{2}
\mid S2 \rangle \langle 2 \mid 
\nonumber \\ &&
+ \sqrt{2} \mid 2 \rangle \langle D2 \mid + 2 \mid S2 \rangle \langle D2 \mid.
\end{eqnarray}
A simple basis change has converted the coherences between empty paths into the only path presence of the photon, while canceling out the presence of the particle in path $\mid 1 \rangle$ by converting it into a coherence between the empty paths $\mid S2 \rangle$ and $\mid D2 \rangle$. A consistent description of particle propagation from $\mid \psi \rangle$ to $\mid f \rangle$ is only possible if the effects of such coherences between empty ports are taken into account, explaining the deterministic delocalization of the particle on its way through the interferometer.

\section{Conclusions}

Quantum interference describes the deterministic relation between different measurement contexts by describing the measurement outcomes of one context as superpositions of measurement outcomes in another context. This creates logical problems, because the combination of initial state $\mid \psi \rangle$ with a specific measurement outcome $\mid f \rangle$ sometimes appears to eliminate all possible intermediate states. The analysis presented here shows that this elimination of intermediate states does not include an elimination of the associated quantum coherences. When the situation is visualized in a three-path interferometer, it is comparatively easy to see that these coherences between empty ports cause an intermediate delocalization of each photon, where a super-localization in both the $\mid 1 \rangle$ and the $\mid 2 \rangle$ paths is made possible by a negative presence in the remaining $\mid 3 \rangle$ path. The expression of contextuality by weak values thus provides an interesting perspective on the role of quantum coherences in the relation between different measurement contexts. It may well be possible to develop a more complete description of quantum mechanics based on a more detailed analysis of the manner in which future measurements modify the quantum coherences defined by the input state.

%%\vfill

\end{document}